\documentclass[prl,twocolumn,amsmath,showpacs]{revtex4}
\usepackage{graphicx}
\usepackage{color}

\begin{document}
\title{Point-contact study of the LuNi$_2$B$_2$C borocarbide superconducting film}
\author{O. E. Kvitnitskaya, Yu. G. Naidyuk, I. K. Yanson}
\affiliation{B. Verkin Institute for Low Temperature Physics and
Engineering, National Academy  of Sciences of Ukraine,  47 Lenin
Ave., 61103, Kharkiv, Ukraine}
\author{T. Niemeier, G. Fuchs,  B. Holzapfel and L. Schultz}
\affiliation{Leibniz-Institut f\"ur Festk\"orper- und
Werkstofforschung Dresden e.V., Postfach 270116, D-01171 Dresden,
Germany}
\date{\today}

\begin{abstract}
We present point-contact (PC) Andreev-reflection measurements of a
superconducting epitaxial $c$-axis oriented nickel borocarbide
film LuNi$_2$B$_2$C (T$_c$=15.9\,K). The averaged value of the
superconducting gap is found to be  $\Delta \simeq$
2.6$\pm$0.2\,meV in the one-gap approach, whereas the two-gap
approach results in  $\Delta_1 \simeq$ 2.14$\pm$0.36\,meV and
$\Delta_2 \simeq$ 3$\pm$0.27\,meV. The better fit of the
Andreev-reflection spectra for the LuNi$_2$B$_2$C--Cu PC obtained
by the two-gap approach provides evidence for multiband
superconductivity in LuNi$_2$B$_2$C. For the first time, PC
electron-phonon interaction (EPI) spectra have been measured for
this compound. They demonstrate pronounced phonon maximum at
8.5$\pm$0.4\,meV and a second shallow one at 15.8$\pm$0.6\,meV.
The electron-phonon coupling constant $\lambda$ estimated from the
PC EPI spectra turned out to be small (with $\lambda_{\rm
{PC}}\sim$ 0.1), like in other superconducting rare-earth nickel
borocarbides. Possible reasons for this are discussed.

\end{abstract}

\maketitle
\section{Introduction}

Among the rare-earth nickel borocarbide superconducting
$Re$Ni$_2$B$_2$C family, the system with $Re=$\,Lu  belongs to the
nonmagnetic ones with the highest T$_{\rm c}$ of about 16\,K.
Although numerous experiments were undertaken to study the
superconducting state there, still there is space for more
detailed investigation. This is most probably due to the
complicated band structure in this compound what may certainly
lead to the anisotropy of the superconducting gap or/and
multi-band (gap) superconductivity. Point-contact (PC) Andreev
reflection spectroscopy is a direct tool to clarify the
superconducting gap characteristics, i.e. its value and its
anisotropy as well as the temperature and magnetic field
dependencies of the gap. Furthermore, PC spectroscopy itself
provides a straightforward information as to the PC
electron-phonon interaction (EPI) function $\alpha^2F(\omega)$
\cite{Naid}, which can be a test for the phonon-mediated
superconductivity.

Several PC studies were performed on single crystals of
LuNi$_2$B$_2$C \cite{Bobrov,Greene}. These experiments have shown
an anisotropy of the superconducting order parameter (gap). In
Ref.\,\cite{Bobrov}, the authors claimed that the data are in
favor of a two-gap model. \textcolor{red}{A similar conclusion was
made also by a study of single crystals and films of the related
compound YNi$_2$B$_2$C in \cite{Bashlak,Bashlakov,raychaudhuri}.
}In this study, we present PC spectroscopy data for epitaxial
c-axis oriented LuNi$_2$B$_2$C films in order to compare these
data with the results obtained for LuNi$_2$B$_2$C single crystals
and to receive information about EPI in LuNi$_2$B$_2$C.

\section{Experimental details}
High-quality LuNi$_2$B$_2$C epitaxial c-axis oriented films with
T$_{\rm c}=15.9$\,K have been fabricated using pulsed laser
deposition. The details of the film preparation are described in
\cite{Niemeier}. XRD measurements show almost perfect c-axis
texture. The films show good homogeneity, very high in-plane and
out-of-plane order and high enough for the thin films residual
resistivity ratios up to 17 at film thicknesses of 250\,nm. The
quality of these films compared to single crystals was tested
using a sample in the as-grown state for B$_{\rm c 2}$
measurements. Upper critical field values in the $<$001$>$
crystallographic direction are higher than those in single
crystals (B$_{{\rm c}2}\simeq 8$T \cite{Shulga}) and show a less
pronounced S-shape behavior.

The PCs were established along the c-direction by the standard
"needle-anvil" method \cite{Naid} touching of the film surface by
a sharpened Cu wire. The differential resistance $dV/dI(V)$ and
the second derivative $d^2V/dI^2(V)$ were recorded by sweeping the
dc current $I$ on which a small ac current $i$ was superimposed
using the standard lock-in technique. The measurements were
performing in the temperature range $T$=1.5-20\,K and in magnetic
fields up to 9\,T.

\subsection{PCS of superconducting energy gap}
Spectroscopic information about superconducting energy gap is
available in the case if the contact diameter $d$ is smaller than
the inelastic electron mean-free path as well as than the
coherence length $\xi$(0).  LuNi$_2$B$_2$C is characterized at low
temperatures by coherence length  about $\xi\approx$ 6\,nm
\cite{Shulga} and elastic electron mean-free path $l \simeq$
10\,nm evaluated from the mentioned below $\rho l$ value.
Estimation of the contact size from its resistance $R_{\rm {PC}}$
using the Wexler formula \cite{Naid}
\begin{equation}
R_{\rm {PC}}\simeq (16\rho l)/3 \pi d^2)+ \rho/2d,
 \label{Wexler}
\end{equation}
gives $d\leq $ 10\,nm for the typical PC resistance $R_{\rm
{PC}}$=10\,$\Omega$ with $\rho \simeq$ 2.7$\mu\Omega$cm and $\rho
l \simeq 3.6 \cdot 10^{–12} \Omega \cdot$ cm$^2$ \cite{bhatnagar}.
Thus, all the mentioned lengths are of the same order of
magnitude, therefore it is not possible to say a priori whether
investigated PCs are in the ballistic (or diffusive), in other
words, in the spectroscopic regime. Independent of the PC
resistance (of course the higher the resistance the higher
probability to be in spectroscopic regime), each PC spectra should
be tested in order to display the spectroscopic features both in
the normal (phonons) and superconducting (gap minima) state.

\begin{figure} [t]
\begin{center}
\includegraphics[width=8cm,angle=0]{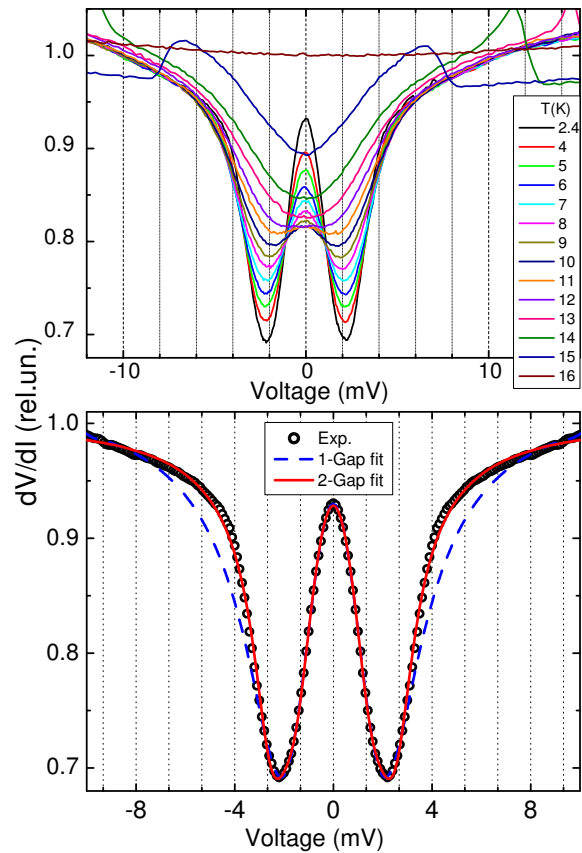}
\end{center}
\vspace{0cm} \caption[] {Upper panel: $dV/dI$ curves of a
LuNi$_2$B$_2$C - Cu contact ($R$ = 6.3\,$\Omega$) established
along the c-axis for varying temperature.  Bottom  panel:
symmetrized and normalized to the normal state $dV/dI$ curve at
$T$ = 2.4\,K (points) together with calculated curves according to
the generalized BTK theory: dashed and solid lines - one-gap and
two-gap fits, respectively.} \label{dvdi}
\end{figure}

We were able to obtain PC $dV/dI$ characteristics, which
demonstrate clear Andreev-reflection (gap) structures --
pronounced minima at $V\simeq \pm\Delta$ at $T<$T$_{\rm c}$ as it
is shown in Fig.\,\ref{dvdi}. For the investigated PCs the
temperature of the vanishing of the superconducting main minimum
in $dV/dI$ was close to 16\,K, that is, close to T$_{\rm c}$ of
the LuNi$_2$B$_2$C film, testifying that superconductivity in the
PCs is not degraded. To retrieve the superconducting gap value
$\Delta$ and other parameters from the Andreev-reflection spectra
the generalized Blonder-Tinkham-Klapwijk (BTK) theory \cite{BTK}
is commonly used. The application of the one-gap BTK model to the
$dV/dI$ curves from Fig.\,\ref{dvdi}(bottom panel) results in a
moderate fit to the experimental curves. Furthermore, the
parameter $\Gamma$, which implies a finite lifetime of carriers
due to inelastic scattering of charge carriers was found to be
rather high (about half of the $\Delta$ value) for this PC. Such
high $\Gamma$ as compared to  $\Delta$ can be connected with the
anisotropy of the superconducting gap or a possible two-gap
superconductivity. The two-gap(band) model is supported by a
recent three-dimensional study of the Fermi surface of
LuNi$_{2}$B$_{2}$C \cite{Dugdale}, where contribution to the
density-of-states (DoS) at the Fermi energy from 3 bands equal
0.24\%, 22.64\% and 77.1\%, respectively, was found. That is, two
bands basically contribute to DoS.

\begin{figure} [t]
\vspace{3cm}
\begin{center}
\includegraphics[width=8cm,angle=0]{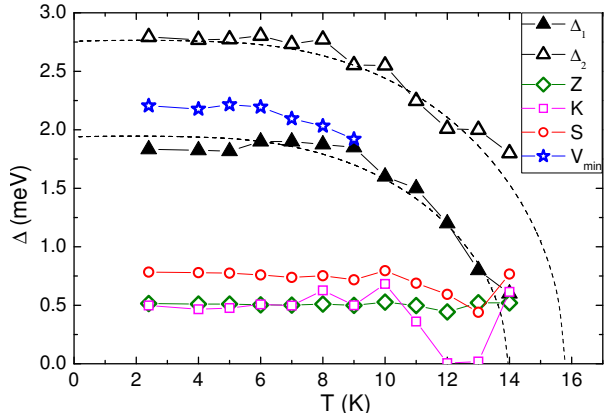}
\vspace{-3cm}
\end{center}
\vspace{0cm} \caption[] {Temperature dependencies of the fitting
parameters: superconducting gaps $\Delta_1$, $\Delta_2$
(triangles), barrier parameter Z (diamonds), contribution (weight
factor) $K$ of $\Delta_1$ (squares), scaling parameter $S$
(snowflakes), position of minima in $dV/dI$ (stars) for the PC
from Fig.\,\ref{dvdi}. The broadening parameter $\Gamma$ is equal
to zero. The dotted lines represent the BCS-like gap behavior.}
\label{DGZ}
\end{figure}

As we can see from Fig.\,\ref{dvdi}(bottom panel) the fitting
within the two-gap model gives a much better agreement with the
experimental $dV/dI$ data. Applying the two-gap fit the
temperature dependencies of the superconducting gaps $\Delta$, the
smearing parameter $\Gamma$ and the barrier strength $Z$ were
established (see Fig.\,\ref{DGZ}). The so-called scaling parameter
$S$ corresponding to the ratio of the experimental $dV/dI$
intensity to the calculated one reflects the quality of the fit
and was tried to keep constant and close to 1 at fitting $dV/dI$.
It turned out that the contribution $K$ (or weight factor) of both
gaps to the spectra is nearly equal $K$=0.5 and the fitting is
possible with $\Gamma$=0. The latter gives strong support for the
used two-gap model which confirms the multiband superconducting
state in this compound. We should also note, that with increasing
temperature the double minimum structure of $dV/dI$ smears out
above 11\,K and this gives more space for the fitting parameters.
Therefore their values are less precise by approaching T$_c$.

The data for the two superconducting gaps presented in
Fig.\,\ref{DGZ} well agree with results reported for
LuNi$_2$B$_2$C single crystals for the c-direction \cite{Bobrov}.
The averaged gap value of $\Delta \simeq$ 2.6$\pm$0.2\,meV for all
contacts (about 30) calculated in the one-band approach turned out
to be close to the averaged gap of $\Delta \simeq$ 2.4--2.5\,meV
reported for PCs on single crystals for the c-direction
\cite{Bobrov,Greene}.Thus, the quality of the investigated films
is comparable to that of the best single crystals. The mean gap
values $\Delta_1 \simeq$ 2.14$\pm$0.36\,meV and $\Delta_2 \simeq$
3$\pm$0.27\,meV established for about 15 PCs from the two-gap
approach are also in line with data obtained for LuNi$_2$B$_2$C
single crystals \cite{Bobrov}. The characteristic values of the
fitting parameters for the measured contacts are presented  in
Table I both for the one- as well for the two-gap approach.

\begin{table}
\caption[]{Average, minimal and maximal values of the
superconducting gap $\Delta$, the "smearing" parameter $\Gamma$
and the "barrier" parameter $Z$ retrieved from the one- and
two-gap approach. }

\begin{tabular}{|c|c|c|c|c|c|c|c|c|}
  \hline

   & $\Delta$(meV) & $\Gamma$(meV) & Z & 2$\Delta$/kT$_c$\\
  \hline
\textbf{Average} & \textbf{ 2.6$\pm$0.2} & \textbf{ 0.54$\pm$0.2} & \textbf{ 0.48$\pm$0.06} & \textbf{ 3.8$\pm$0.3} \\
    Minimal & 2.12 & 0.22 & 0.32 & 3.21 \\
     Maximal & 2.85 & 0.9 & 0.57 & 4.16 \\
  \hline
\end{tabular}
\vspace{1cm}

\begin{tabular}{|c|c|c|c|c|c|}
  \hline

   & $\Delta_1$(meV) & $\Gamma_1$(meV) &$\Delta_2$(meV) & $\Gamma_2$(meV)& Z\\
  \hline
\textbf{Average} & \textbf{ 2.14$\pm$0.36} & \textbf{ 0.36} & \textbf{ 3$\pm$0.27} &\textbf{ 0.25} & \textbf{ 0.47} \\
    Minimal & 1.65 & 0 & 2.6 & 0 & 0.32  \\
     Maximal & 2.6 & 0.74 & 3.45 & 0.73 & 0.54  \\
  \hline
\end{tabular}
\end{table}

\subsection{PCS of quasiparticle excitations}

\begin{figure} [b]
\vspace{3cm}
\begin{center}
\includegraphics[width=8.5cm,angle=0]{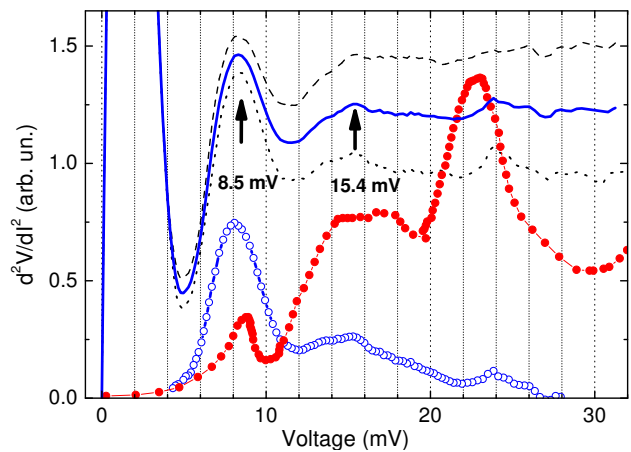}
\vspace{-3cm}
\end{center}
\caption[] { Averaged for both polarities \textcolor{red}{second
harmonic signal} $V_2\propto d^2V/dI^2$ (solid curve) measured in
magnetic field of 7\,T for the PC from Fig.\,\ref{dvdi} taken at
$T$ = 2.4\,K in comparison with the phonon DOS for LuNi$_2$B$_2$C
\cite{Gompf} (solid circles). The open circles show $d^2V/dI^2$
with subtracted background in the form similar to $\log(V)$ above
6\,mV \textcolor{red}{(see also Fig.\,\ref{md2vdi2})}. Dashed and
dotted curves show raw data of $d^2V/dI^2$ for this contact for
two bias polarity.} \label{d2vdi2}
\end{figure}

As it was mentioned in the introduction, the PC spectroscopy makes
it possible to study the electron-phonon interaction (EPI). The
second derivative of the $I(V)$ curve of the ballistic contact at
low temperatures is directly proportional to the PC EPI function
$\alpha^2F(\omega)$ \cite{Naid}. The latter can be expressed using
measurable signals as
\begin{equation}
\alpha^2F(\omega) = \frac{3}{2\surd2}\frac{\hbar v_F}{ed}\frac{V_2}{V_1^2},
\label{EPIF}
\end{equation}
where $e$ is the electron charge, $d$ is the PC diameter, $v_F$ is
the Fermi velocity, $V_1$ and $V_2$ are the rms amplitude of the
first and the second harmonics of the modulating signal
respectively, which are proportional to the first d$V$/d$I$ and
the second d$^2V$/d$I^2$ derivatives of the $I(V)$ curve of the
investigated PC, respectively.

\begin{figure} [t]
\begin{center}
\includegraphics[width=7.5cm,angle=0]{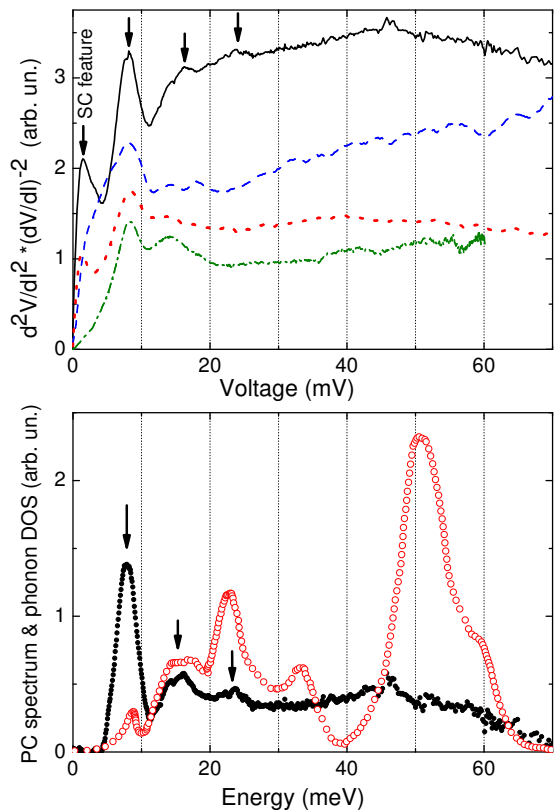}
\end{center}
\caption[] { Upper panel: Reduced second harmonic signal
$V_2/V_1^2$ [$\propto d^2V/dI^2(dV/dI)^{-2}=R^{-1}dR/dV$, where
$R=dV/dI$ ] measured in magnetic field of 9\,T for different PCs
with $R$ = 3.7, 5.6, 4.7 and 2.9\,$\Omega$ (from top to bottom)
taken at $T$ = 2.2 - 4.2\,K. \textcolor{red}{Long-dash line shows
tentative background behavior for the upper curve in the form
$a\,\log(b\,V)+c$.} Bottom panel: PC spectrum (solid circles) with
subtracted background for the contact with $R$ = 3.7\,$\Omega$
along with the phonon DOS for LuNi$_2$B$_2$C \cite{Gompf} (open
circles). } \label{md2vdi2}
\end{figure}

In Fig.\,\ref{d2vdi2}, the $d^2V/dI^2$ curves of the PC from
Fig.\,\ref{dvdi} are shown. We applied the magnetic field to
suppress the superconductivity in the PC, because the magnitude of
the superconducting features in the PC spectra near zero bias is
much larger than that of the maxima caused by the EPI. As we can
see from Fig.\,\ref{d2vdi2}, the magnetic field of 7\,T was not
high enough to suppress the superconductivity completely (B$_{\rm
c2} \approx$ 9\,T at 2\,K \cite{Niemeier}) and the huge feature at
bias about 2\,mV is due to the residual superconductivity. Besides
the feature due to the superconducting gap, a clear-cut maximum
slightly  above 8\,mV and a more smeared one at around 15\,mV are
well distinguished (see Fig.\,\ref{d2vdi2}). The mentioned peaks
correspond to the low energy phonon maxima in the phonon DOS of
LuNi$_2$B$_2$C \cite{Gompf} (see Fig.\,\ref{d2vdi2}, symbols).
Whereas the high energy part of the obtained PC spectra contains
no visible features. In Fig.\,\ref{md2vdi2}, we present a set of
$d^2V/dI^2$ curves for different PCs measured up to bias voltages
of about 80\,mV. By analyzing all of obtained PC spectra (about
30) with the visible phonon features, we have found, that the
$d^2V/dI^2$ curves of the LuNi$_2$B$_2$C PCs display phonon maxima
at 8.5$\pm$0.4\,mV and 15.5$\pm$1.0\,mV (averaged for 8 PC)
similarly to the PC from Fig.\,\ref{d2vdi2}. In contrast they do
not contain contributions from the other phonon peaks at 23, 33,
and 50\,mV observed in the phonon DOS of LuNi$_2$B$_2$C (shown in
Fig.\,\ref{md2vdi2} by the circle symbols). Only the PC with $R =
3.7\,\Omega$ in Fig.\,\ref{md2vdi2} shows a weak hump at 23\,mV.
The reason for the absence of contribution at high-energy phonon
maxima could be a deviation from the ballistic (spectroscopic)
regime at higher voltages due to increase of the EPI and a
shortening of the inelastic mean free path of electrons.
\textcolor{red}{It is worth to mention that in
\cite{Dervenagas,Stassis} phonon softening of two branches was
observed below T$_c$. In our case, we did not observe these modes
around 4-5\,meV, probably because we have measured PC EPI spectra
by suppressing the superconducting state, while in other case a
huge gap maximum makes it impossible to see any other features
below 5-6\,meV (see, e.g., Fig.\,3, where the superconducting
state is not fully suppressed). Another interesting question
arises, whether it is possible to separate the contribution of
each (two) bands to the PC EPI spectrum? The problem here is
similar to the analogous one raised in the tunneling spectroscopy
in \cite{Dolgov}, where the authors concluded that it is not
possible to obtain several band splitted EPI functions from a
single function of the tunnel current.}

After subtraction of the background from the measured PC spectra
we established the EPI function according to (2) (using maximal
$v_F = 3.6\cdot10^7$ cm/s calculated for LuNi$_2$B$_2$C
\cite{bhatnagar}) and estimated the EPI parameter
$\lambda=2\int\alpha^2F(\omega) \omega^{-1}d\omega$. The latter
even for the spectra with the maximal intensity was found to be
not bigger than 0.1. This value is an order of magnitude smaller
than the $\lambda$ values between 0.5 and 1 from dHvA data for the
$c$ - direction of LuNi$_2$B$_2$C \cite{Bergk,Isshiki} or $\lambda
\approx 0.5-0.8$ estimated from STM measurements on LuNi$_2$B$_2$C
\cite{Martinez}. Such small values of $\lambda$ could be due to
some simplifications of the PC spectroscopy theory where only the
free electron model and a single band Fermi surface are used. Some
issues of $\lambda$  evaluation from PC spectra were discussed in
\cite{Naidyuk}. Most of all, the coupling constants estimated from
PC spectra should be considered as lower bonds for the coupling
constant $\lambda$ relevant for superconductivity. Here we also
would like to mention that the discussed $\lambda$ parameter is
some kind of transport EPI constant and in general its value is
different from the Eliashberg EPI constant (see Table 3.1 in
\cite{Naid}), but the difference by one order of magnitude is of
course confusing. \textcolor{red}{The calculation of the PC EPI
function may shed light on this issue. It will allow to separate
the bands contributions, to estimate the integral intensity of the
spectrum and to determine the relative contribution of each phonon
branch to the PC EPI spectrum. Sure, it would be a very helpful,
but also a sophisticated task and it is beyond the scope of this
experimental paper.}

Another reason for a reduced intensity of the measured PC spectra
might be the elastic scattering, which can be larger in the PC
core than in the bulk sample due to less perfect surface
properties and stresses at the PC formation. As follows from the
PC spectroscopy theory, the magnitude of nonlinearity in PC
spectra is proportional to $l/d$ \cite{Kulik} in the diffusive
regime $l\ll d$. Fig.\,5 shows the intensity of the main peak in
the PC spectra, which indeed shows strong scattering in the range
of about one order. Since we do not see any trend in the intensity
of the main peak vs PC resistance which related to the PC size
(diameter), we expect that the spectra with the higher intensity
are in (or close to) the ballistic regime. Thus, deviation from
the ballistic regime of the current flow in the investigated
contacts can not be the reason of small $\lambda$ values, which
were calculated for the PCs with the maximal intensity.

Similar low $\lambda$ values as we found for LuNi$_2$B$_2$C were
obtained also from the PC spectra of YNi$_2$B$_2$C, HoNi$_2$B$_2$C
\cite{Bashlakov, Naidyuk, Yanson} and recently for TmNi$_2$B$_2$C
(not yet published). There is also general similarity of PC
spectra for the mentioned compounds which are characterized by a
prevailing first phonon maximum (excluding CEF peaks in
HoNi$_2$B$_2$C and TmNi$_2$B$_2$C). As mentioned above, by
determining $\lambda$ from PC spectra one can underestimates the
superconducting EPI. In this context, a notable large
$\lambda_{\rm {PC}}$=0.85 was reported for one special PC spectra
of  DyNi$_2$B$_2$C, while $\lambda_{\rm {PC}}$= 0.25 was found for
more typical PC spectra of this compound \cite{BobrovDy}. However,
there by calculating $\lambda_{\rm {PC}}$ the low energy intensive
"magnetic" peak at $\sim$ 5\,mV was also taken into account, which
gives about half of the $\lambda$  value. Therefore, the
concerning EPI contribution to $\lambda$ in DyNi$_2$B$_2$C is
again close to 0.1, if we eliminate the mentioned extra high
$\lambda$  value and the "magnetic" peak contribution to
$\lambda$.

\begin{figure}
\vspace{3cm}
\begin{center}
\includegraphics[width=8cm,angle=0]{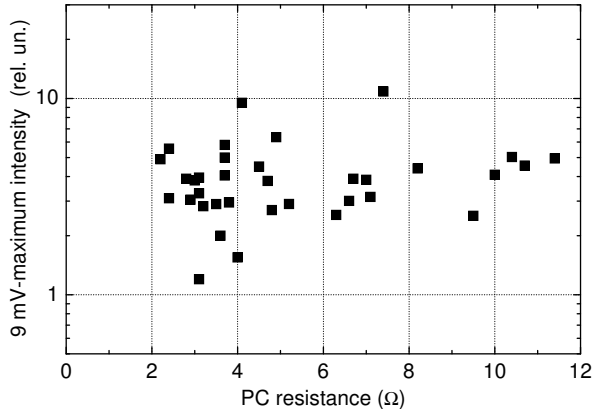}
\end{center}
\vspace{-3cm} \caption[] {\textcolor{red}{Scatter in the}
intensity of the first phonon maximum in $d^2V/dI^2$ curves for
all measured LuNi$_2$B$_2$C PCs versus their resistance.}
\label{intens}
\end{figure}

\section{Conclusion}
We investigated the superconducting energy gap and EPI in
LuNi$_2$B$_2$C using an epitaxial c-axis oriented film by PC
spectroscopy. The mean value of the superconducting gap is found
to be about 2.6$\pm$0.2\,meV (2$\Delta$/kT$_c=3.8\pm0.3$) in the
one-gap model what is very close to the values reported for PC
measurements on single crystals \cite{Bobrov,Greene}. However, the
fitting of $dV/dI$ curves favors the two-gap approach which
provides strong support for the multiband superconducting state in
this compound. For the two-gap approach the averaged gap values
are found to be $\Delta_1 \simeq$ 2.14$\pm$0.36\,meV and $\Delta_2
\simeq$ 3$\pm$0.27\,meV.

For the first time for LuNi$_2$B$_2$C, we succeeded to measure EPI
PC spectra  with the distinct phonon peaks at 8.5$\pm$0.4\,mV and
15.8$\pm$0.6\,meV. Therefore, we can conclude that these low
energy phonons play a preferential role in the pairing mechanism.
The EPI spectra are in general similar to those measured for other
nickel borocarbides \cite{Bashlakov,Naidyuk,Yanson,BobrovDy}
showing predominance of the first phonon peak in EPI  for all
these compounds.

\section*{Acknowledgements}
Two of the authors (O.E.K., Yu.G.N.) would like to thank the
Alexander von Humboldt Foundation for the support, Dr. S.-L.
Drechsler for the useful discussions and Dr. K. Nenkov for the
technical assistance.


\end{document}